%% file: causalityandnetworks.tex
\documentclass[12pt, a4paper]{article}
\usepackage[english]{babel}
\usepackage[T1]{fontenc}
\usepackage[scaled]{helvet}
\renewcommand*\familydefault{\sfdefault}
\usepackage{eurosym}
\usepackage[dvips]{graphics}
\usepackage{amsmath, amssymb}
\usepackage[usenames,dvipsnames,svgnames,table]{xcolor}
\usepackage{subfig}
\usepackage{array,booktabs}
\usepackage{times}
\usepackage{multicol}
\usepackage{afterpage}
\usepackage{xargs} 
\usepackage{ifthen}
\usepackage{pifont}
\usepackage{setspace}
\usepackage{stmaryrd}
\usepackage{wrapfig,framed}
\usepackage{floatrow} 
\usepackage{caption}
\usepackage{mathabx}
\usepackage{enumitem} 
\usepackage{colortbl} 
\usepackage{lipsum}
\usepackage{microtype}                                                          
\usepackage[cm]{fullpage}
\usepackage{ulem} 
\renewcommand{\em}{\it}
\usepackage{url}
\usepackage{hyperref} 
\usepackage{geometry}
\usepackage{fancyhdr}
\usepackage{footnote} 
\usepackage[super,nomove]{cite}

\DeclareColorBox{shaded}{\colorbox{yellow!20}}
\addto\captionsenglish{%
}

\makeatletter
\renewcommand\section{\@startsection{section}{2}{-0.9cm}%
	{0.4cm }%
	{0.3cm }%
	{\reset@font\large\bfseries}}
\makeatother

\newcommand{\figref}[1]{\hyperref[#1]{Fig.\ref{#1}}}

\definecolor{aenlever}{RGB}{100,75,100}
\definecolor{newcol}{RGB}{0,75,100}
\colorlet{cfond}{black!5}
\colorlet{newcolC}{newcol}
\colorlet{newcolL}{ForestGreen}
\colorlet{citecolorSAFE}{ForestGreen}
\colorlet{citecolor}{ForestGreen}
\colorlet{newcolU}{newcol}
\definecolor{crimson}{RGB}{220,20,60}
\colorlet{mycitecolor}{crimson}
\colorlet{mathcolor}{gray}
\definecolor{turquoise4}{RGB}{0,134,139}
\definecolor{cadetblue4}{RGB}{83,134,139}
\definecolor{tan4}{RGB}{139,90,43}
\definecolor{red4}{RGB}{139,0,0} 
\definecolor{indianred}{RGB}{176,23,31}
\definecolor{crimson}{RGB}{220,20,60}
\definecolor{goldenrod1}{RGB}{255,193,37}
\definecolor{royalblue4}{RGB}{39,64,139}
\definecolor{azure4}{RGB}{131,139,139}
\definecolor{azure5}{RGB}{105,111,111}
\definecolor{darkolivegreen}{RGB}{85,107,47}
\definecolor{emeraldgreen}{RGB}{0,201,87}

\hypersetup{colorlinks=true, linkcolor=Black!90,urlcolor=OrangeRed,
  citecolor=citecolor}

\newcommand{\citetodo}[1]{{\color{red}\bf [??]}}

\newcommand{\citeal}[1]{{\color{citecolor}\hspace{-2pt}\textsuperscript{[}\cite{#1}\textsuperscript{]}}}

\newcommand{\citemycite}[2]{{\color{citecolor}\textsuperscript{[}\cite{#1}{\color{mycitecolor}\colorlet{citecolor}{mycitecolor}\textsuperscript{\citepunct}
\cite{#2}\colorlet{citecolor}{citecolorSAFE}}\textsuperscript{]}}}

\newcommand{\mycite}[1]{{\color{mycitecolor}\colorlet{citecolor}{mycitecolor}\textsuperscript{[}\cite{#1}\textsuperscript{]} \colorlet{citecolor}{citecolorSAFE}}}
\newcommand{\myciteOL}[1]{{\color{mycitecolor}\colorlet{citecolor}{mycitecolor}[\citenum{#1}]\colorlet{citecolor}{citecolorSAFE}}}

\newcommand{\myurl}[1]{\scalebox{0.8}[1.1]{\textls[-50]{\url{#1}}}}

\usepackage{multibib}
\newcites{latex}{\LaTeX-Literature}
\usepackage{bibentry}

\RequirePackage{tikz} 
\usetikzlibrary{snakes,backgrounds,arrows,matrix,petri,shapes} 
\RequirePackage{extarrows}

\RequirePackage{fourier-orns}

\def\ie{{\it i.e.}~} 
\def\vs{{\it vs}~} 
\def\eg{{\it e.g.}~}
\def\cf{{\it cf.}~}

\newcommand{\Ai}[0]{\scalebox{0.4}{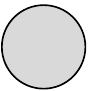}}
\newcommand{\Aj}[0]{\scalebox{0.4}{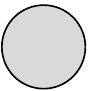}}
\newcommand{\Aone}[0]{\scalebox{0.4}{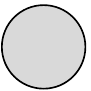}}
\newcommand{\Atwo}[0]{\scalebox{0.4}{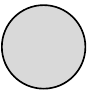}}
\newcommand{\Athree}[0]{\scalebox{0.4}{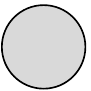}}
\newcommand{\Afour}[0]{\scalebox{0.4}{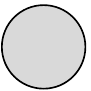}}
\newcommand{\Afive}[0]{\scalebox{0.4}{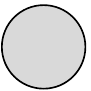}}
\newcommand{\Asix}[0]{\scalebox{0.4}{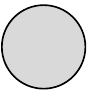}}
\def\N{{\ensuremath {\cal N}}}

\def\B{\ensuremath{\mathbb{B}}}
\def\Bn{\ensuremath{\B^n}}

\newcommand{\set}[1]{\ensuremath{\{#1\}}}

\newcounter{listitem}
\renewcommand{\thelistitem}{(\alph{listitem})~}
\newcommand{\litem}{{\color{black!60}\thelistitem}\refstepcounter{listitem}}

\newcounter{listitemr}
\newcounter{listitems}
\newcounter{listitemM}
\newcounter{listitema}
\renewcommand{\thelistitema}{(\arabic{listitema})~}
\newcommand{\litema}{{\color{black!60}\thelistitema}\refstepcounter{listitema}}

\def\endlitema{\setcounter{listitema}{1}}
\def\endlitem{\setcounter{listitem}{1}}

\newcounter{observation}
\newcounter{observationG}

\begin{document}

\title{{\Large Causality  and Networks}} 
\author{{\large Mathilde Noual}}
\maketitle


Causality is omnipresent in scientists' verbalisations of their understanding,
even though we have no formal consensual scientific definition for it.  In
Automata Networks, it suffices to say that automata ``{influence}'' one another
to introduce a notion of causality.
One might argue that this merely is an incidental side effect of preferring
statements expressed in natural languages to mathematical formulae.
The discussion of this paper shows that if this is the case, then it is
worth considering the effects of those preferences on the contents of the
statements we make and the formulae we derive. And if it is not the case, 
then  causality  must  be  worth some scientific attention {\em
  per se}. \linebreak
In any case, the paper illustrates how the innate sense of causality we have
may be made deliberate and formal use of without having to pin down the elusive
notion of causality to anything fixed and formal that wouldn't do justice to the
wide range of ways it is involved in science-making.

\section{Boolean Automata Networks}

A network is a set of entities/parameters causing each other to undergo
change.\bigskip

A Boolean Automata Network (BAN) $\N=\set{f_i:\Bn\to
  \B=\set{0,1}\,|\,i\in V=\llbracket 1,n\rrbracket}$ is a set of Boolean functions  that define the way automata in the set
$V=\llbracket 1,n\rrbracket$ interact with one another.The interaction graph
$G=(V,A)$ of $\N$ represents these interactions. 
Arc $(i,j)\in A$ belongs to the interaction graph $G$ of $\N$ if and only
if $x_i$ appears in the Conjunctive Normal Form of $f_j(x)$ -- \ie if and only
if $f_j(x)$ depends on $x_i$. 
When $\N$ is in state
$x=(x_1,x_2,\ldots,x_j,\ldots,x_n)\in \Bn$, automaton $j\in V$ has the possibility of undergoing the change of
states $x_j\leadsto \neg x_j$ if it is unstable,  meaning: if $j\in U(x)=\set{i\in V:
  f_i(x)\neq x_i}$.
It only actually does undergo the change of states $x_j\leadsto \neg x_j$ if on
top of being unstable in $x$, $j$ is also {\em updated} in $x$ (by us).
\bigskip\bigskip

As mentioned above, in BANs, it suffices to say that automata
``{influence}'' one another to introduce a notion of causality.  It is
very natural to interpret arc $(i,j)\in A$ as meaning ``automaton $i$ can cause
automaton $j$ to change states''. 
There are other notable examples of causality's spontaneous involvement in the
BAN formalism\ldots

\section*{Additional definitions}

Before moving on to the  next sections, I will introduce here some additional
definitions that the sequel refers to.

\subsection*{The parallel update schedule}

The {parallel update schedule} of a BAN $\N=\set{f_i:\Bn\to \B=\set{0,1}\,|\,i\in
  V=\llbracket 1,n\rrbracket}$ is defined by the function $F:(x_1,x_2,\ldots,
x_n)\in \Bn\mapsto (f_1(x),f_2(x),\ldots,f_n(x))\in Bn$. The {\em transition
  graph} induced by the parallel update schedule is the graph of function $F$,
namely $T_p=(\Bn,\rightarrowtriangle\hspace{-0.8em}\blacktriangleright)$ where
$(x,y)\in \rightarrowtriangle\hspace{-0.8em}\blacktriangleright \iff y=F(x)$.\medskip

In $x\in Bn$, the parallel update schedule exploits all instabilities of $U(x)$,
\ie{} it exploits all possibility of synchronous updates. 

\subsection*{The asynchronous setting}

Asynchronism is not an update schedule. It is a notion that can mean
$|U(x)|=1$.\linebreak It however usually is rather taken to refer to a very
popular setting of the literature of (B)ANs. In this setting, the transition
graph considered is $T_a=(\Bn,\rightarrowtriangle)$, namely the {\em
  asynchronous transition graph} where $(x,y)\in \rightarrowtriangle \iff
\exists i\in U(x): D(x,y)=\set{i\in V:x_i\neq y_i}=\set{i}$. \medskip

In the asynchronous setting, all sequential trajectories are assumed to be
possible. \linebreak All trajectories involving some synchronous updates are assumed not to
be.

\subsection*{General transition graphs}

The general transition graph of a BAN is the graph $T_g=(\Bn,\to)$ where the
relation $\to$ is given by: $(x,y)\in \to\iff\forall i\in V, x_i\neq y_i:\, i\in
U(x)$.
\medskip

General transition graphs are simply state transition systems. They were
introduced in \myciteOL{GIGS} to to study the {\em behavioural possibilities} of
networks as opposed to the {\em dynamics}.

\subsection*{Block-sequential update schedules}

A {\em block-sequential update schedule} (BSUS) is an update schedule that
updates automata of a network in a certain deterministic periodic order and with a certain
amount of synchronism in the updates (possibly none). Within a period of updates, all automata
are updated exactly once. \medskip

Formally it can be defined as a function $\nu: A \to \set{-1,+1}$ such that
starting in $x\in\Bn$ at the beginning of the period, $\forall (i,j)\in A$,
$\nu(i,j)=-1$ iff $i$ is updated strictly before $j$ is.\linebreak Thus, when $j$ is
updated, $i$ is already in state $f_i(x)$. Otherwise, if $\nu(i,j)=+1$, then $i$
is updated no sooner than $j$ is so when $j$ is updated, $i$ is still in state
$x_i$.\linebreak\vspace{-4mm}

The ``degree of synchronism'' of $\nu$ might be formally matched to the number of
arcs $(i,j)\in A$ such that $\nu(i,j)=+1$. This is only a suggestion however
because the definition of BSUS given here is not common.\label{BSUS}\medskip

\begin{figure}[h!]
\begin{tabular}{c@{\hspace{15mm}}c}
\fcolorbox{black}{black!5}{\scalebox{0.65}{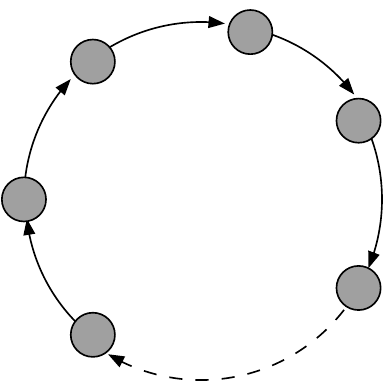}} \\[-25mm] &
$\begin{cases} f_i: x\mapsto x_{i-1} ~~~ \forall i\neq 0\\ f_1:x\mapsto \neg
  x_n\end{cases}$\\[15mm]
\end{tabular}
\caption{A negative {\em Boolean Automata Cycle}, a.k.a. negative {\em
    feedback loop}. {\sc left} Its interaction graph. {\sc right:} Its defining
  local update functions.  The BAC owes its negativity  to the odd
  number of negative arcs in the interaction graph.}
\label{fig:oscillations}
\label{fig:cycle}
\end{figure}

 Traditionally, BSUSs are rather defined as lists of disjoint blocks of automata
 $(B_k)_{k\leq K}$ where $\biguplus_{k\leq K}B_k=V$. Automata in a block are
 updated in parallel. The blocks are updated sequentially
 \citeal{Robert1986B,Aracena2009}. With this definition, a BSUS has more
 synchronism if it has less and larger blocks.
For example consider the BAN of \figref{fig:cycle}. It is what we call a Boolean
Automata Cycle (BAC). One particular BSUS of this BAC is the following which
defines $K=n$ blocks of size $1$: $\set{n},\set{n-1},\ldots,
\set{3},\set{2},\set{1}$. This BSUS sequentially updates automata in the reverse
order of the cycle. It has no synchronism.  Another BSUS of the BAC is:
$\set{6,\ldots, n-1,n},\set{2,3,\ldots, 5},\set{1}$ which defines $K=3$
blocks. Two of them are of size greater than $1$ so this BSUS {\em does} have
some synchronism.\linebreak It updates at once all automata in block
$B_1=\set{6,\ldots, n-1,n}$. Then it updates at once all automata in block
$B_2=\set{2,3,\ldots, 5}$. And only then does it update automaton
$1$. \linebreak The BSUS $\set{2,3,\ldots,n-1,n},\set{1}$ has even more
synchronism since it updates at once all automata except automaton $1$, before
it updates automaton $1$.\medskip

Notably, all three of these BSUSs are equivalent to $\nu:A\to
\set{-1,+1}$ defined by $\nu(i,j)=1 \Leftrightarrow (i,j)\neq (n,1)$. 

\section{Synchronism \vs Precedence}\label{sec:precedence}

In some circumstances,
a synchronous updating of automata seems to cause local instabilities to linger
in the network longer than they might with more sequentiality.
In other terms, synchronism seems to inhibit the decrease of the number $|U(x)|$
of automata that are unstable. And as a consequence, it also seems to be
responsible for the global asymptotic instability of the BAN
\citemycite{Elena2009T}{AUTOMATA2010,AAM2011,GIGS}.
\bigskip

Indeed, on the one hand, the BSUSs of BACs that allow for the most local and
global instability are the BSUSs with a less \mbox{synchronism
  \citemycite{Elena2009T}{AUTOMATA2010}.} Moreover, we know that with the
parallel update schedule a BAC has many attractors. And the majority of those
attractors involve several local instabilities circulating through the network
\mycite{DAM2010}.
On the other hand, in the asynchronous setting, all BAC trajectories lead to an
attractor that has either $0$ or $1$ instabilities \citeal{Remy2003}.
And the general transition graphs of BACs show that at least some sequentiality is required
at some point  to reduce the number of instabilities in the BAC.
\bigskip

All this supports the idea  that   synchronism is responsible for entertaining instabilities.
In communities focusing on asynchronous Automata Networks  to model 
genetic regulation networks, synchronism is actually notorious for this effect.
Synchronism's alleged tendency to  artificially entertain instabilities in a
network is a very common informal argument invoked to undermine the realism of
updating schedules involving synchronism, and the relevance of research that
doesn't rule synchronism out altogether.
As I was cutting my teeth on such research
\mycite{stageM1,stageM2,IJMS2009,PLOS2010,DAM2010,AUTOMATA2010,JTB2011,AAM2011,thesis},
it was necessary for me to investigate the causal link between synchronism and
instability \mycite{mysensy}.\bigskip

\begin{figure}[h!]
\begin{tabular}{c@{\hspace{15mm}}c}
\fcolorbox{black}{black!5}{\scalebox{0.6}{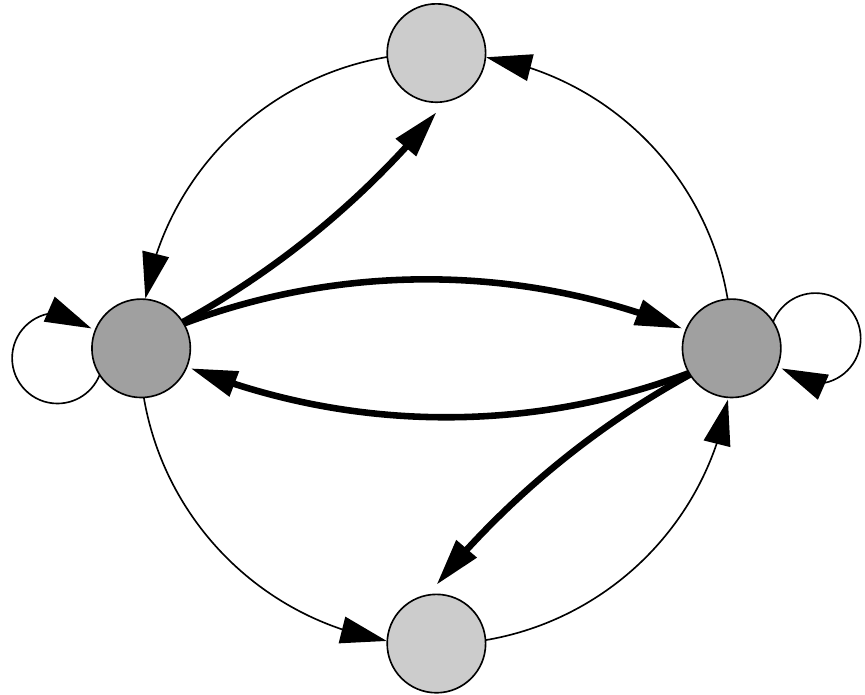}} \\[-40mm]
& $\begin{cases}f_0:x\mapsto x_2\vee(x_0\wedge\neg
         x_1)\\f_1:x\mapsto x_3\vee(\neg x_0\wedge x_1)\\ f_2:x\mapsto\neg x_0\wedge
         x_1\\ f_3:x\mapsto  x_0\wedge\neg x_1\end{cases} $\\[25mm]\end{tabular}
\caption{This is the BAN given in \protect \myciteOL{mysensy} as an example of
  ``synchronism-sensitive'' BAN. Its asynchronous transition graph $T_a$ is represented and discussed in \figref{fig:contrexGT}
  together with the addition of some synchronism to $T_a$.
}
\label{fig:precedence}
\label{fig:contrex}
\end{figure}

As it turns out, when a small amount of synchronism is added to an otherwise
asynchronous BAN, synchronism either has no lasting effect (most frequent case
by far), or it has some (this is the case with the BAN of
\figref{fig:contrex}). And if it has some then, precisely, its effect is to
stabilise all local instabilities. And in that, it stabilises the whole BAN
\mycite{mysensy} (see \figref{fig:contrexGT}).\bigskip

\begin{figure}[h!]
\tikz[scale=1] {
\filldraw[fill=cyan, draw=cyan, ] (-0.5,-0.7)--(0,-0.7)--(0,-0.2)--cycle;
\filldraw[fill=Yellow, draw=Yellow,]  (0.5,-0.7)--(0,-0.7)--(0,-0.2)--cycle;
\node (0) at (0,0)  {0000};
\node (1) at (5,-10)  {0001};
\node (2) at (-5,-10)  {0010};
\node (3) at (0,-16)  {0011};
\node (4) at (-2,-4)  {0100};
\node (5) at (2,-10)  {0101};
\node (6) at (-2,-6)  {0110};
\node (7) at (2,-12)  {0111};
\node (8) at (2,-4)  {1000};
\node (9) at (2,-6)  {1001};
\node (10) at (-2,-10)  {1010};
\node (11) at (-2,-12)  {1011};
\node (12) at (0,-2)  {1100};
\node (13) at (2,-8)  {1101};
\node (14) at (-2,-8)  {1110};
\node (15) at (0,-14)  {1111};
\draw[-,line width=2mm,left,cyan]  (-0.1,-1.7) to node  {{\scriptsize 0}} (-0.1,-0.4);
\draw[-,line width=2mm,right,Yellow]  (0.1,-1.7) to node  {{\scriptsize 1}} (0.1,-0.4);
\draw[-triangle 90,right,cyan]  (12) to node {{\scriptsize 0}} (4);
\draw[-triangle 90,right,Yellow]  (12) to node {{\scriptsize 1}} (8);
\draw[-triangle 90,right,ForestGreen]  (4) to node  {{\scriptsize 2}} (6);
\draw[-triangle 90,right,RawSienna]  (8) to node  {{\scriptsize 3}} (9);
\draw[-triangle 90,right,cyan]  (6) to node  {{\scriptsize 0}} (14);
\draw[-triangle 90,right,Yellow]  (9) to node  {{\scriptsize 1}} (13);
\draw[-triangle 90,right,Yellow]  (14) to node {{\scriptsize 1}} (10);
\draw[-triangle 90,right,cyan]  (13) to node {{\scriptsize 0}} (5);
\draw[-triangle 90,right,RawSienna]  (10) to node {{\scriptsize 3}} (11);
\draw[-triangle 90,right,ForestGreen]  (5)  to node {{\scriptsize 2}} (7);
\draw[-triangle 90,right,Yellow]  (11) to node {{\scriptsize 1}} (15);
\draw[-triangle 90,right,cyan]  (7)  to node {{\scriptsize 0}} (15);
\draw[-,right,cyan]  (-0.6,-15.9)  to node {} (-2,-15.9);
\draw[-triangle 90,right,cyan] (-2,-15.9)  to node {{\scriptsize 0}} (11);
\draw[-,right,Yellow]  (0.6,-15.9)  to node {} (2,-15.9);
\draw[-triangle 90,right,Yellow] (2,-15.9)  to node {{\scriptsize 1}} (7);
\draw[-,right,ForestGreen]  (3)  to node {} (5,-16);
\draw[-triangle 90,right,ForestGreen] (5,-16)  to node {{\scriptsize 2}} (1);
\draw[-,right,RawSienna]  (3)  to node {} (-5,-16);
\draw[-triangle 90,right,RawSienna]  (-5,-16)  to node {{\scriptsize 3}} (2);
\draw[-,right,ForestGreen]  (2)  to node {{\scriptsize 2}} (-5,0) ;
\draw[-triangle 90,right,ForestGreen] (-5,0)  to node  {} (0) ;
\draw[-,right,RawSienna]  (1)  to node {{\scriptsize 3}} (5,0);
\draw[-triangle 90,right,RawSienna] (5,0)  to node {} (0);
\draw[-triangle 90,above,cyan]  (2)  to node {{\scriptsize 0}} (10);
\draw[-triangle 90,above,Yellow]  (1)  to node {{\scriptsize 1}} (5);
\draw[-triangle 90,left,ForestGreen]  (15) to node {{\scriptsize 2}} (13);
\draw[-triangle 90,left,ForestGreen]  (11) to node {{\scriptsize 2}} (9);
\draw[-triangle 90,left,ForestGreen]  (10) to node {{\scriptsize 2}} (8);
\draw[-triangle 90,right,ForestGreen]  (14) to node {{\scriptsize 2}} (12);
\draw[-triangle 90,right,RawSienna]  (15) to node {{\scriptsize 3}} (14);
\draw[-triangle 90,right,RawSienna]  (7) to node {{\scriptsize 3}} (6);
\draw[-triangle 90,right,RawSienna]  (5) to node {{\scriptsize 3}} (4);
\draw[-triangle 90,left,RawSienna]  (13) to node {{\scriptsize 3}} (12);
}
\caption{The asynchronous transition graph of the BAN of \figref{fig:contrex}
  with the additional synchronous transition $1100\stackrel{01}{\longrightarrow}
  0000$. All other synchronous transitions of the BAN do not have a lasting
  effect: they merely shortcut asynchronous trajectories. With a purely asynchronous updating of automata states,
  the BAN has two attractors: \protect \litema  the stable state $0000$ ($U(0000)=V$)
  and \protect \litema\endlitema a large cyclic attractor comprised of the 12
  states at the centre. When the transition $1100\stackrel{01}{\longrightarrow}
  0000$ is added, the BAN looses its second attractor. The thing to notice is
  the following. Consider the state $x=1100$. In this state, if the two
  instabilities -- that of automaton $0\in U(x)$ and that of automaton $1\in
  U(x)$ -- are not settled -- \ie if the synchronous transition is not made --
  then either the two parts of what constitutes a XOR $x_0\oplus x_1$ , namely
  $(x_0\wedge \neg x_1)$ and $(\neg x_0 \wedge x_1)$ start being schlepped
  around in the network, from one automaton to the next, thereby entertaining a
  certain tacit dependency between the instabilities of automata, precisely the
  dependencies that characterises the second attractor of the asynchronous updating. It isn't yet clear
  however whether or not they assemble straightforwardly into $x_0\oplus x_1$ at
  some point. But the BAN of \figref{fig:contrex} being a minimal example of
  synchronism-sensitive BAN \protect \mycite{mysensy}, this suggests there is a
  relation between synchronism-sensitivity and non-monotony.}
\label{fig:contrexGT}
\end{figure}


So on the one hand, synchronism seems to cause instabilities to linger, on the
other it `causes' instabilities to disappear.
The quotes around 'causes' are to help keep in mind that the notion of causality
we are manipulating here is not one that is formally defined.  
In both situations, synchronism is seen as a cause of something, but in none is
it actually the logical implicant of anything.
%
Causality, which is not logical implication, is much more charged with meaning
than logical implication is. Here, it is charged in particular with the specific
meaning of the term ``synchronism'' that is used in each of the two
situations (\cf paragraph on Block sequential update schedules on Page \pageref{BSUS}).\medskip

The conclusion we can draw from the apparent contradiction is the
following. \linebreak We
have taken a certain perspective. According to this perspective, 'synchronism'
is a meaningful notion: it is a notion that pinpoints actual causes of notable
effects we are interested in. More specifically, according to this perspective,
'synchronism' is the type of thing that can cause an effect on local
instabilities. With this perspective, we run into an apparent
contradiction. {\em Two opposite effects are incumbent on the same cause.}
%
%
%
And that in itself is a call for an update of perspective. Synchronism (and
asynchronism for that matter) might not be the appropriate notion to explain either
of the effects we want it to explain in relation  to instabilities.\bigskip

Nonetheless, the intuitions expressed by the two different cause/effect relations
involving synchronism and instabilities cannot be baseless. Or at least this
much is true: considering that they are baseless is not going to help us make
any progress, we had better trust that they aren't and go looking for their
basis in order to formalise it.
Ruling out synchronism as a possible medium for the effects of entertaining and
settling instabilities means that we need a new candidate cause that can convey a finer form of
causality accounting coherently for both effects.
Comparing the two contradictory situations reveals that {\em precedence} might
be a more reasonable choice for that. Just like synchronism, it is involved in both
situations. And it is a plausible basis for both sides of the contradiction as
it can explain both situations coherently.
Indeed, in the first situation, instead of ``{\em Synchronism causes instabilities to
linger}'' or rather instead of ``{\em In the absence of synchronism, instabilities do
not linger (as much)}'', it seems more relevant to use the following explanation
of what happens to instabilities:

\begin{center} ``{\em Assuming event B owes its
  possibility to event A not having occurred, then, in case of precedence of the
  occurrence of possible event A over the occurrence of possible event B, the
  occurrence of  A causes the impossibility of the occurrence of  B.}''
\end{center}

meaning that the implementation of instability  A (\ie the possibility of A) is
enough to settle at once both
instability A and instability B.
And in the second situation, instead of ``{\em Synchronism settles instabilities}'',
we can use the following explanation: 
\begin{center}
``{\em If two events A and B are possible, if the occurrence of each results in
    the disappearance of one instability, then in the absence of any
  interference of precedence, the occurrence of A and B results in the
  disappearance of two instabilities.}''
\end{center}
 The effects studied in
\myciteOL{mysensy} were classified in terms of varying degrees of {\em
  sensitivity to synchronism}. But following this discussion, {\em sensitivity to precedence of causally related
  events} might be a more accurate and relevant way of coining the same effects.
\bigskip

Generally, if $(i,j)\in A$ and $i\in U(x),\, j\notin U(x)$, then updating
$i$ \underline{before} $j$ -- assuming
$j$ owes its stability in $x$ to $i$ being in state $x_i$ --
causes  $i$ to stabilise and $j$ to become unstable in turn.
If both automata $i\in U(x),\, j\in U(x)$ are unstable in $x$, then --
assuming $x_i$ is the reason for $j$'s instability in $x$ -- updating $i$
\underline{before} $j$ will stabilise $i$ in state $\neg x_i$ and thereby also
cause the stabilisation of $j$ in state $x_j=f_j(\bar{x}^{i})$.
\bigskip

It no longer is a matter of synchronism.\bigskip

On top of putting forward precedence as a more relevant and accurate cause, the
comparison of the two initial contradictory cause/effect relations emphasises
all the other differences there are between the two situations. \medskip

 Determinism of the parallel update schedule and of BSUSs is one of them. And
 this difference with the asynchronous setting relates to the notable fact that
 asynchronism is not an update schedule, and it is thereby not comparable with
 sequential update schedules.
$T_a$, just like $T_g$, defines a state transition system representing {\em
  behavioural possibilities}, {as opposed} to {\em defined dynamical
  behaviours}. It is not clear how to compare what is called an ``attractor'' in
the asynchronous setting (the terminal strongly connected components of $T_a$)
to the attractors of dynamics defined by specific deterministic update
schedules.\medskip

Periodicity -- perhaps even more subtly: the specific kind of redundancy
inherent to periodicity -- is another notable difference between the two
situations.  In the lead of F. Robert \citeal{Robert1986B}, a great many studies
have been supporting the general idea that ``{update schedules have great
  influence on the dynamics of
  BANs}''~\citemycite{Elena2008,Goles2008,Tosic10}{thesis,DAM2011,AAM2011,AUTOMATA2010}.
But just like precedence and determinism, until periodicity's own effects aren't
studied {\em per se}, there is no rigorous way to form a more reliable intuitive
understanding of what generally causes the entertainment of local instabilities
under BSUSs and of what, other than asynchronism, tends to prevent the
entertainment of local instabilities when synchronism is not exploited.


\section{Interlude}\label{sec:interlude}

The previous section discussed a case where two different effects are
intuitively attributed to the same cause in different contexts, and shows how to
learn from that by letting the  following question be raised: 
{\em How is the cause
  really involved in the generation of the effect?}
and letting it lead to the next question, namely {\em  What else is involved? in particular,
  what finer cause might  the original cause be a facade for or an abstraction of?}
\bigskip

Another case we can learn from is when two different causes are intuitively
invoked to explain the same effect in different contexts. In this case the
following question calls for an answer: {\em What is the effect's common
  implicant/generating mechanism?} This case is illustrated in the next section.

\section{Oscillations and the difference between experience of change and communication of change}\label{sec:oscillations}

In some contexts, negative feedback loops, or negative cycles (\cf
\figref{fig:oscillations}) are considered to be directly responsible for the
asymptotic oscillations of BANs
\citeal{Remy2008a,Blanchini000562,Thomas1981,Snoussi1998,Siebert2010}. In
different contexts, BANs can have asymptotic oscillations without negative
cycles \citemycite{Elena2008,Demongeot2003}{DAM2010,LATA2012}.  Asymptotic
oscillations can therefore not intuitively be {attributed} to negative cycles
(nor to the specific context of each case for that matter). Negative cycles are
not the {\em fundamental} generating mechanism of asymptotic oscillations. A
close comparison between those apparently contradictory cases reveals however,
that with or without negative cycles, asymptotic oscillations are essentially
generated the same way: ``something'' is disallowing the collapse of an offset
between the actual state of a certain automaton $i$ and the pending influence it
sends out to itself possibly via other automata.
In the first context, that ``something'' is a negative feedback loop. In the
other contexts, it is a combination of a positive feedback loop, some {\em in
  situ} potentiality (defined by $x\in \Bn$), and some \sout{synchronism}
absence of precedence. In those other contexts the same effect is produced with
what   can be argued to be less numerous and less elaborate means,
requiring neither precedence nor negation, just the feedback loop and {\em in
  situ} information.\bigskip

Now, we have identified a common generating mechanism with two possible
implementations  explaining the same effect in two different contexts. It
remains to address the questions that naturally follow starting with {\it What other ways are
  there to implement this mechanism?} and {\em Can this mechanism be held responsible
  for other effects?}

\section{Interlude}

The way Sections \ref{sec:precedence} and \ref{sec:oscillations} suggest to make
use of the notion of causality is by considering it as essentially progressive
so that we let causal relations point towards the signs that already exist of
their own coming obsolescence.
A particular advantage of taking such a flexible, yet deliberate and rigorous
perspective on causality is that it keeps us aware of the possibility of having
different causes causing the same effects and thus that the explanations we
currently have of the effects we currently are interested in might still be
worth investigating even if they appear to be perfectly functional and complete
explanations. This perspective on causality also prepares us to deal
with new situations in which we observe the same familiar effects in a system,
without having any formal reason to believe that the same implicants we
are used to are responsible.
And it  favours keeping in mind the possibility that causes we have represented
of the effects we have observed, can themselves be abstractions of more subtle,
atomic mechanisms operating at a lower level of abstraction.  It encourages
finding ways to exploit this possibility in order to refine our explanations and
understanding.
Generally, this approach to causality emphasises the fact that for a significant
part, what we manipulate as scientists is representations of the objects that we
study, not just  the objects themselves.  And in emphasising this
fact, it allows us to take advantage of it. 
\bigskip

Now of course, BANs can be studied with purely mathematical interests and
perspectives.  Then, causality is not such an important concern. Implication is
enough.  But because it applies only to specific properties of specific systems
in specific conditions, implication has the downside of being much less portable
than causality.
Moreover, in theory, we are free to pick any restriction on the kinds of BANs we
consider. And often, when BANs are studied for purely mathematical
reasons, the priority is to pick a restriction that will help
derive new mathematical results \ldots

\section{Non-monotony and efficiency}\label{sec:NM}

\begin{figure}[h!]
\begin{tabular}{@{\hspace{0mm}}l@{\hspace{0mm}}l@{\hspace{-3mm}}}
\scalebox{0.85}{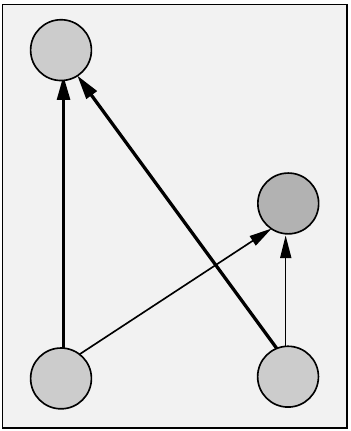}&\scalebox{0.85}{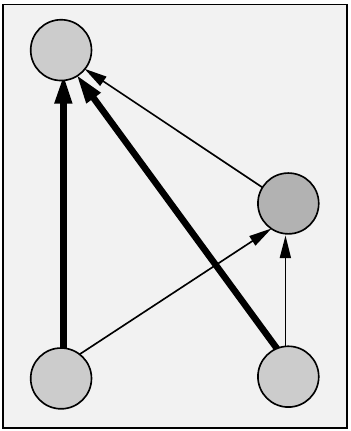}\\
$\begin{cases}
f_1: x\mapsto x_2\oplus x_3\\
f_4:x\mapsto x_2\vee x_3
\end{cases}$ & $\begin{cases} f_1: x\mapsto (\neg x_2\vee \neg x_3)\wedge x_4\\
f_4:x\mapsto x_2\vee x_3
\end{cases}$\\[-3mm]\end{tabular}
\caption{{\sc left:} Non-monotone interactions with automaton $1$. {\sc right:} Monotone
  interactions with automaton $1$ that can produce exactly the same effects if  the
  right relative arrangement in time of automata updates is chosen.}
\label{fig:NMcode}
\end{figure}

\begin{figure}[h!]
\begin{tabular}{@{\hspace{0mm}}l@{\hspace{0mm}}l@{\hspace{-3mm}}}
\scalebox{0.6}{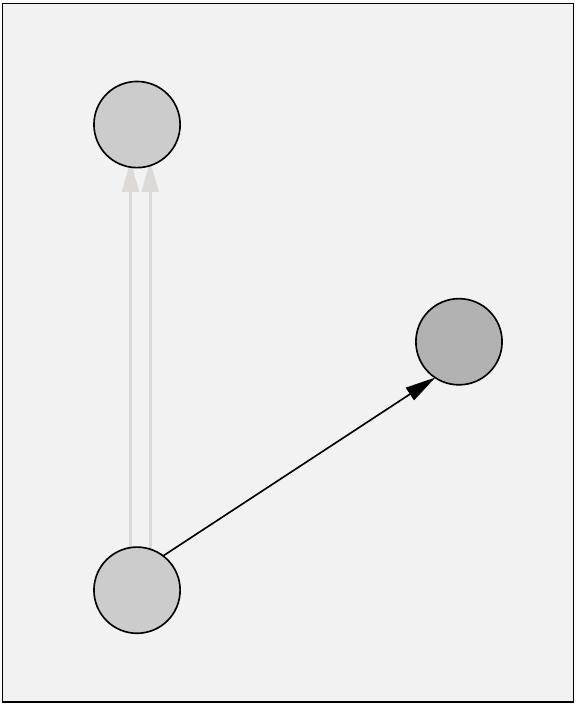}&\scalebox{0.6}{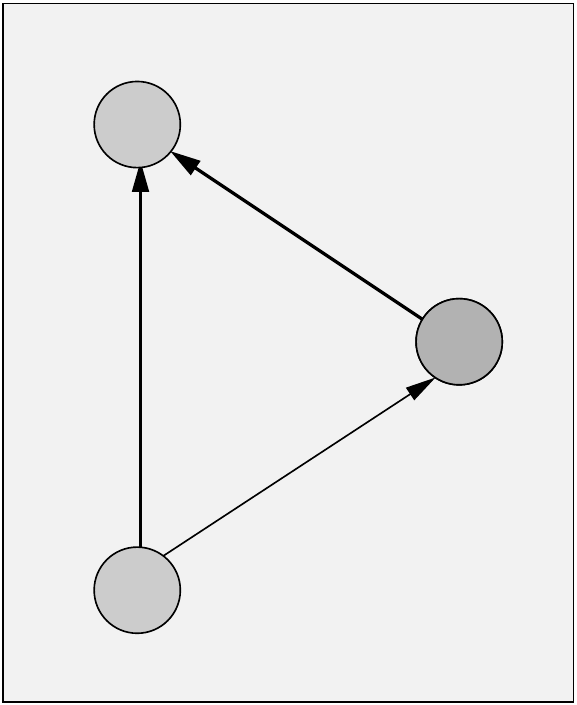}\\
$\begin{cases}
f_i: x\mapsto x_j\oplus   x_j=0\\
f_k: x\mapsto x_j
\end{cases}$ & 
$\begin{cases}
f_i: x\mapsto x_j\oplus x_k\\
f_k: x\mapsto x_j
\end{cases}$ \\[-3mm]\end{tabular}
\caption{{\sc left} and {\sc right},  the function defining how automaton $i$
  behaves is  $f_i(x)=x_j\oplus x_\ell$ for $\ell \in \set{j,k}$. 
Whether or not $j$ is considered to be an influencer of $i$ depends on
  whether $\ell=j$ or $\ell \neq k$. And yet, the non-monotone
  BAN on the right can be made to behave exactly as the monotone BAN on the left by
  imposing the systematic precedence of the update of $k$ over that of $i$ so
  that $\forall x,\,f_i(x)=x_j\oplus f_k(x)=xj\oplus x_j$. }
\label{fig:NM1}
\end{figure}

A common restriction motivated by biological considerations is the restriction
to monotone (B)ANs
\citeal{Remy2003,Chaouiya2004,Aracena200649,Mendoza1998,Blanchini000562}. 
\figref{fig:NMcode} shows that  a monotone BAN can behave just like a
non-monotone BAN in some conditions  (see also  \myciteOL{Persp}).
Conversely, \figref{fig:NM1} shows that  a non-monotone BAN can behave just like a
monotone BAN in some conditions.
Studying monotone BANs and non-monotone BANs is therefore not enough. Even
comparing them is not enough because at most what we get from doing that is a
list of monotone BAN properties that non-monotone BANs don't share, and a list
of non-monotone BAN properties that monotone BANs don't have.
If a monotone BAN can behave exactly like non-monotone BAN and {\it vice-versa}, then
what we need to know, is how {\em non-monotony} and {\em monotony} work: {\em
  What are the effects/properties each of them is strictly responsible
  for?}\bigskip

As illustrated in \figref{fig:NMcode}, for a monotone BAN's behaviour to look
just like that of a non-monotone BAN, it requires the right arrangement in time
of automata updates, and it requires us ignoring certain {\em intermediary
  steps}. For the monotone BAN on the right of \figref{fig:NMcode} to behave
like the non-monotone BAN on the left, automaton $4$ needs to be updated
systematically before automaton $1$ is, and the two sequential transitions
$x=(x_1,x_2,x_3,x_4)\stackrel{{\color{gray!70}\scriptsize
    4}}{\longrightarrow}x'=(x_1,x_2,x_3,f_4(x))\stackrel{{\color{gray!70}\scriptsize
    1}}{\longrightarrow}x''=(f_1(x'),x_2,x_3,f_4(x))$ need to be summed up into
just one transition $x=(x_1,x_2,x_3,x_4)\longrightarrow
x''=(f_1(x'),x_2,x_3,f_4(x))$. Incidentally, since automaton $4$ no longer has
any impact, we can  loose it altogether and concentrate on the asynchronous
transition $(x_1,x_2,x_3)\longrightarrow (f_1(x'),x_2,x_3)$.  In any case, the
example of  \figref{fig:NMcode} shows that
the intrinsic effects of non-monotony that cannot be reproduced
without non-monotone $f_i$'s, \ie effects that represent the real difference
between the two types of BANs, are strongly related to time flow (the relative
order of events considered) and to the degree of precision of our observations
of global trajectories (with more or less intermediaries).
This suggests that non-monotony {\em per se} might impact in terms of the efficiency of the
execution of mechanisms in  BANs.
\bigskip

Of course, such a claim needs to find some formal support. Even though
the notion of causality it conveys is unlikely to be exhausted in one proven
mathematical statement, the claim might nonetheless be supported to some extent
by putting forward proven mathematical statements that are compatible with it
just like the example of \figref{fig:NMcode} is.
But \figref{fig:NM1} suggests that just like with synchronism, at some point, we
might need to call into question non-monotony's capacity at explaining
relevantly the effects we attribute the responsibility of to it. And together
with that, we'll have to call into question the perspective that places the
focus on the distinction monotony/non-monotony rather than on a more subtle
distinction, one that can provide more complete explanations with less means. In
other terms, non-monotony will eventually have to give way to a finer
explanation of the same effects. And considering \figref{fig:NM1}, it seems that
two notions implied by non-monotony might then be put forward instead of
non-monotony: a notion of ``witness'' and a notion of ``inconsistent parallel
transfer of the same information''.

\section{Non-expansivity}\label{sec:NE}

In some (asynchronous) contexts \citeal{Remy2008a,AracenaRS16}, a BAN $\N$ is
defined using the global update function $F:x\mapsto (f_1(x),\ldots,f_n(x))~=\N$
which happens to coincide with the function defining the dynamics of $\N$ under
a parallel update schedule.
The definition is equivalent to the one chosen in Section 1, namely
$\N=\set{f_1,\ldots,f_n}$. But it makes it more natural to pick restrictions on
BANs that are given by properties of $F$. An example is
non-expansivity: $\forall x,y\in \Bn:\,$\mbox{$|\set{x_i\neq y_i}|\geq
  |\set{F(x)_i\neq F(y)_i}|$}.
Intuitively, $F$'s non-expansivity corresponds to the BAN having a form of
global instantaneous potential. Assuming $F$'s non-expansivity happens to favour
the derivation of some results about asynchronous BANs
\citemycite{Richard2013expasynch}{BMB2013}. But precisely, asynchronous BANs are
BANs whose dynamical constraints forbid them the use of this global potential.
This makes it difficult to grasp intuitively the meaning of the results in
question. And mathematical results that are more difficult to have an intuitive
grasp on are often results that are more difficult to generalise and relate to
other results.

\section{Assumptions and intuitions}

The examples of sections \ref{sec:NM} and \ref{sec:NE} show that however mathematically sound are the mathematical
results we prove thanks to mathematical assumptions/restrictions, disregard for
intuitive causality at worst stakes the applicability of those results, and
generally limits our progression.
In particular, without deliberate care, there is no reason to believe that we
owe the deriving of these results to some deeper opportune relevance of the
mathematical assumptions/restrictions. There is no reason to believe there is
anything in the assumptions/restrictions that could enable the generalisation
of the results beyond the setting they define, nor anything that could at least
guarantee the relatability of the results to other existing results.
The primary reason why we might have managed to derive anything under a
particular assumption/restriction might be that it is an extremely strong assumption/restriction.
It might be like studying crows by concentrating on the class of crows that a
human being has reported seeing picking up a piece of pink plastic wrapper. It
might be quite unclear what it is that we are studying and learning about
exactly: the original (mathematical) object of interest? the restriction?
And in the case of Automata Networks, this means that the only hope to actually
build a global understanding of networks lies in the platonic wager that it will
necessarily ``emerge''\label{causality-emerge} from the accumulation of
independent studies made of particular models of networks in different settings,
sometimes juxtaposed for comparison.
In lines with Section \ref{sec:precedence}, let us mention as an example that a
great amount of rigorous theoretical attention has been invested in the study of
a large variety of update schedules
\citemycite{Robert1986B,Snoussi1989,ConfuseParallelAndSynchronism3,Mortveit2001,Reidys2006,Goles1990B,Elena2008,Goles2008,Aracena2009,Tosic10,Tournier2005T}{thesis,DAM2011,AAM2011,AUTOMATA2010}. And
yet, none of the studies it occasioned, not even the ones that {\em compare}
specific update schedules, have yet provided explicit answers to the kind of
fundamental questions mentioned in Section \ref{sec:interlude}, including
questions about the effect {\it per se} of the redundancy inherent to the
periodicity of a periodic update schedule, the effect {\it per se} of the
determinism of a deterministic update schedule, the effect {\it per se} of
varying amounts of instabilities, and of varying degrees of imposed precedence
between their exploitation.
This can however naturally be made up for by  sometimes \litem focusing
primarily on general -- \ie{} fundamental and portable -- network attributes
(\eg synchronism, non-monotony, reversibility, subsequence) rather than on
network-specific properties (\eg particular interaction digraphs, particular
$f_i$'s) or restrictions, and by \litem\endlitem studying rigorously the
involvement of those attributes in the behavioural possibilities of
networks (as opposed to describing formally a specific  dynamics). This way, what we know of networks and what we don't can be
re-examined and clarified.

\section{Time flow and potential storage/computation space}

The systems we consider as scientists are often assumed to be conditional to
properties of time (flow). Despite this, their formal definitions sometimes
allows properties of time flow to mingle or overlap with their own properties
(\cf remark made in \myciteOL{Persp}, Section 2, about update constraints and the
dynamical systems view on BANs).  Thus, time flow contributes to the way we
attribute properties to a system.  Yet, it isn't always clearly distinguished
from causality.\medskip

Generally, when it isn't regarded as a pre-existing constraint on the systems we
consider, time flow -- or rather just ``Time'' -- is seen as a ``resource'', suggesting
that {we have a tacit obligation to use it sparingly, and, without fail, in
  {finite} quantity}. Whatever we call it, we tend to assume that it
pre-exists both the systems we study and the attention we invest in them, and
that it frames both the systems' behavioural possibilities and the leeway we take on
them.
Operational Research works its satisfaction and optimisation problems around it;
Bio-informatics builds models out of what it knows of it or despite
what it doesn't \citeal{OnTimedModelsofGeneNetworks,Klarner2011,Comet2010,Siebert2006a,ThomasK2001};
 and even for Concurrency, time flow is  mostly
something {\em in} which  distributed pieces of 
computation might be reunited  \citeal{BookOfTraces,ConcurrentSystems}.
Despite its very dense presence in the scientific landscape, time flow is seldom
a {\em primary} object of our attention.
This results in plenty space for spontaneous interpretations to operate. And in
particular it gives carte blanche to a natural tendency we have to expediently
distinguish, confuse, overlook or classify issues and properties related to time
flow such as simultaneity, synchronicity, precedence, subsequence, difference,
determinism, periodicity, causality, time scale, change, process of change,
realism, duration.
As a result, we    miss out on \litem the {vicariance} of some of these
properties for which time flow might actually not be the exclusive medium, and
\litem possible leeway through this vicariance.
\bigskip

\clearpage
In an isolated network, time flow is precisely determined by the set of all
possible events considered in the network and by their relative
order\footnote{Note how different this is from saying that in a network, all
  possible events considered happen {\em in} Time and their order is defined
  {\em by} Time or relatively to it. In particular, it informally implies  a
  different sense of the term ``isolated''.}.\bigskip

Giving to causality the attention advocated for in the previous sections leads to
putting emphasis on the notion of time flow. It requires to
systematically isolate the involvement time flow has in the effects we take
interest in, and decompose it into more atomic properties that can more tightly
be held responsible for the effects.
Thus, the effects of specific properties of time flow (possibly implicitly
assumed by the specific framework, \eg precedence of certain event occurrences
over certain others) may be compared with effects of properties specific to
networks (properties of the $f_i$s). And thus, we can work towards a better
understanding of network sensitivity to time flow. We can start clarifying
the kind of information time flow is a vehicle of, as well as the part of the
information encoded in the clockworks of networks (the $f_i$'s) that can equivalently
be encoded into time flow.\medskip

  Eventually we may also clarify the kind of
computation complexity time flow can manage. Indeed, results \mycite{mysensy} suggest
that under very specific conditions, attributes of time flow might participate
in the overall network computation in ways that are comparable to logical
gates.\bigskip

As argued in \myciteOL{Persp}, the notion of synchronism in BANs is often
confused with a notion of simultaneity in 'reality' referring to dates
represented by the same point on  \mbox{The Time-Line \mycite{Persp}.}
Nonetheless, in BANs, the possibility of synchronism is an atemporal relation
between possible events. It relates events of the set $\set{x_i\leadsto \neg
  x_i\,|\, i\in U(x)}$. Events that are synchronously possible are independently
caused: they can't be the cause of one another since they haven't yet occurred. So
synchronism implies nothing about time flow. Precisely, the occurrences of the
events it relates are not ordered relatively to one another -- or at least, in
the case of asynchronous BANs, not {\em yet} in $x$.\linebreak
If BANs are supposed to model what we observe of real systems, then the
implications of this are noteworthy. {\em How the occurrences of synchronously
  possible events actually end up arranging themselves relatively to one
  another} is information. And it is information that is {\em not} modelled.
We can either claim that we have the information (\cf assumptions mentioned on
page 9 of \myciteOL{Persp}), or we can look for it. If we claim we have the
information, then as \myciteOL{Persp} demonstrates, we still cannot model it with
BANs: we can only {\em interpret} a feature of BANs as representing it which is
very different.
To explain a particular ordering of occurrences of synchronously
possible events requires certain care in how the notion of causality is handled.
The model (\ie{} the BAN) accounts for the synchronous possibility of the
events, and not the relative arrangement of their occurrences.
If the model represents what we know of the real system at a certain level of
abstraction, then the information we are looking for -- about the relative
ordering of synchronously possible events -- cannot be found at the same level
of abstraction, or else, we did a bad job in modelling the real system and/or
in choosing a formalism to do that. Building a model with the capacity of revealing causal
relationships that are otherwise not apparent to us, is not the same thing as
building a representation  of the set of effects we are interested in.
The relative ordering of the occurrences of synchronously possible events
creates new relationships between those events -- relationships that otherwise
don't exist as evidenced by the synchronous possibility of these events. Thus,
it brings together otherwise presently independent pieces of information and can
cause them to interact as they wouldn't have if they had been left
independent. In the case of the BAN of \figref{fig:contrex}, the effect of
adding this relation evokes that of logical connectors  ($\wedge,\vee,\oplus$,
\cf caption of
\figref{fig:contrexGT}). Of course, again, this claim begs to be investigated.
\bigskip


\section{Conclusion}

There is not much reason to believe that there is a scientific definition of the
notion of causality that can be found to account completely for the way
causality serves science-making. Centuries of modern
science-making haven't been enough to find anything close to a satisfying
proposition.  The notion is  much too large and diverse to be fitted
exhaustively into a fixed predefined formalisation of it.
Besides,  causality stands for humans' {instinctive} way of grasping the
world. It is an essential part of our motivation to explore the world further
and have more of it grasped.  It makes sense to have it serve science-making in an
unformalised intuitive way.  This is however not a reason for letting it serve
science-making in a fortuitous way.
And we have shown that rather than sidelining the manifestations of this
intuitive instinct of ours, we can advantageously supervise its interference
with the scientific formalism we use, and let it serve as a pointer towards
knowledge in need of further explicit formalisation.
\bigskip

\renewcommand{\refname}{Bibliography}
\bibliographystyle{unsrt}
{\small \bibliography{bibliofr}}

\end{document}

%% file: automatoni.pdf_t
\begin{picture}(0,0)%
\includegraphics{automatoni.pdf}%
\end{picture}%
\setlength{\unitlength}{4144sp}%
\begingroup\makeatletter\ifx\SetFigFont\undefined%
\gdef\SetFigFont#1#2#3#4#5{%
  \reset@font\fontsize{#1}{#2pt}%
  \fontfamily{#3}\fontseries{#4}\fontshape{#5}%
  \selectfont}%
\fi\endgroup%
\begin{picture}(398,398)(2097,-935)
\put(2206,-826){\makebox(0,0)[lb]{\smash{{\SetFigFont{20}{24.0}{\familydefault}{\mddefault}{\updefault}{\color[rgb]{0,0,0}$i$}%
}}}}
\end{picture}%

%% file: automatonj.pdf_t
\begin{picture}(0,0)%
\includegraphics{automatonj.pdf}%
\end{picture}%
\setlength{\unitlength}{4144sp}%
\begingroup\makeatletter\ifx\SetFigFont\undefined%
\gdef\SetFigFont#1#2#3#4#5{%
  \reset@font\fontsize{#1}{#2pt}%
  \fontfamily{#3}\fontseries{#4}\fontshape{#5}%
  \selectfont}%
\fi\endgroup%
\begin{picture}(398,398)(2097,-935)
\put(2206,-826){\makebox(0,0)[lb]{\smash{{\SetFigFont{20}{24.0}{\familydefault}{\mddefault}{\updefault}{\color[rgb]{0,0,0}$j$}%
}}}}
\end{picture}%

%% file: automaton1.pdf_t
\begin{picture}(0,0)%
\includegraphics{automaton1.pdf}%
\end{picture}%
\setlength{\unitlength}{4144sp}%
\begingroup\makeatletter\ifx\SetFigFont\undefined%
\gdef\SetFigFont#1#2#3#4#5{%
  \reset@font\fontsize{#1}{#2pt}%
  \fontfamily{#3}\fontseries{#4}\fontshape{#5}%
  \selectfont}%
\fi\endgroup%
\begin{picture}(398,398)(2097,-935)
\put(2206,-826){\makebox(0,0)[lb]{\smash{{\SetFigFont{20}{24.0}{\familydefault}{\mddefault}{\updefault}{\color[rgb]{0,0,0}1}%
}}}}
\end{picture}%

%% file: automaton2.pdf_t
\begin{picture}(0,0)%
\includegraphics{automaton2.pdf}%
\end{picture}%
\setlength{\unitlength}{4144sp}%
\begingroup\makeatletter\ifx\SetFigFont\undefined%
\gdef\SetFigFont#1#2#3#4#5{%
  \reset@font\fontsize{#1}{#2pt}%
  \fontfamily{#3}\fontseries{#4}\fontshape{#5}%
  \selectfont}%
\fi\endgroup%
\begin{picture}(398,398)(2097,-935)
\put(2206,-826){\makebox(0,0)[lb]{\smash{{\SetFigFont{20}{24.0}{\familydefault}{\mddefault}{\updefault}{\color[rgb]{0,0,0}2}%
}}}}
\end{picture}%

%% file: automaton3.pdf_t
\begin{picture}(0,0)%
\includegraphics{automaton3.pdf}%
\end{picture}%
\setlength{\unitlength}{4144sp}%
\begingroup\makeatletter\ifx\SetFigFont\undefined%
\gdef\SetFigFont#1#2#3#4#5{%
  \reset@font\fontsize{#1}{#2pt}%
  \fontfamily{#3}\fontseries{#4}\fontshape{#5}%
  \selectfont}%
\fi\endgroup%
\begin{picture}(398,398)(2097,-935)
\put(2206,-826){\makebox(0,0)[lb]{\smash{{\SetFigFont{20}{24.0}{\familydefault}{\mddefault}{\updefault}{\color[rgb]{0,0,0}3}%
}}}}
\end{picture}%

%% file: automaton4.pdf_t
\begin{picture}(0,0)%
\includegraphics{automaton4.pdf}%
\end{picture}%
\setlength{\unitlength}{4144sp}%
\begingroup\makeatletter\ifx\SetFigFont\undefined%
\gdef\SetFigFont#1#2#3#4#5{%
  \reset@font\fontsize{#1}{#2pt}%
  \fontfamily{#3}\fontseries{#4}\fontshape{#5}%
  \selectfont}%
\fi\endgroup%
\begin{picture}(398,398)(2097,-935)
\put(2206,-826){\makebox(0,0)[lb]{\smash{{\SetFigFont{20}{24.0}{\familydefault}{\mddefault}{\updefault}{\color[rgb]{0,0,0}4}%
}}}}
\end{picture}%

%% file: automaton5.pdf_t
\begin{picture}(0,0)%
\includegraphics{automaton5.pdf}%
\end{picture}%
\setlength{\unitlength}{4144sp}%
\begingroup\makeatletter\ifx\SetFigFont\undefined%
\gdef\SetFigFont#1#2#3#4#5{%
  \reset@font\fontsize{#1}{#2pt}%
  \fontfamily{#3}\fontseries{#4}\fontshape{#5}%
  \selectfont}%
\fi\endgroup%
\begin{picture}(398,398)(2097,-935)
\put(2206,-826){\makebox(0,0)[lb]{\smash{{\SetFigFont{20}{24.0}{\familydefault}{\mddefault}{\updefault}{\color[rgb]{0,0,0}5}%
}}}}
\end{picture}%

%% file: automaton6.pdf_t
\begin{picture}(0,0)%
\includegraphics{automaton6.pdf}%
\end{picture}%
\setlength{\unitlength}{4144sp}%
\begingroup\makeatletter\ifx\SetFigFont\undefined%
\gdef\SetFigFont#1#2#3#4#5{%
  \reset@font\fontsize{#1}{#2pt}%
  \fontfamily{#3}\fontseries{#4}\fontshape{#5}%
  \selectfont}%
\fi\endgroup%
\begin{picture}(398,398)(2097,-935)
\put(2206,-826){\makebox(0,0)[lb]{\smash{{\SetFigFont{20}{24.0}{\familydefault}{\mddefault}{\updefault}{\color[rgb]{0,0,0}6}%
}}}}
\end{picture}%

%% file: cycle.pdf_t
\begin{picture}(0,0)%
\includegraphics{cycle.pdf}%
\end{picture}%
\setlength{\unitlength}{4144sp}%
\begingroup\makeatletter\ifx\SetFigFont\undefined%
\gdef\SetFigFont#1#2#3#4#5{%
  \reset@font\fontsize{#1}{#2pt}%
  \fontfamily{#3}\fontseries{#4}\fontshape{#5}%
  \selectfont}%
\fi\endgroup%
\begin{picture}(1755,1746)(3402,-2245)
\put(5041,-1456){\makebox(0,0)[lb]{\smash{{\SetFigFont{10}{12.0}{\rmdefault}{\mddefault}{\updefault}{\color[rgb]{0,0,0}\colorbox{black!5}{$+$}}%
}}}}
\put(3466,-1141){\makebox(0,0)[lb]{\smash{{\SetFigFont{10}{12.0}{\rmdefault}{\mddefault}{\updefault}{\color[rgb]{0,0,0}\colorbox{black!5}{$+$}}%
}}}}
\put(3511,-1816){\makebox(0,0)[lb]{\smash{{\SetFigFont{10}{12.0}{\rmdefault}{\mddefault}{\updefault}{\color[rgb]{0,0,0}\colorbox{black!5}{$-$}}%
}}}}
\put(4771,-781){\makebox(0,0)[lb]{\smash{{\SetFigFont{10}{12.0}{\rmdefault}{\mddefault}{\updefault}{\color[rgb]{0,0,0}\colorbox{black!5}{$+$}}%
}}}}
\put(4051,-646){\makebox(0,0)[lb]{\smash{{\SetFigFont{10}{12.0}{\rmdefault}{\mddefault}{\updefault}{\color[rgb]{0,0,0}\colorbox{black!5}{$+$}}%
}}}}
\put(4501,-691){\makebox(0,0)[lb]{\smash{{\SetFigFont{11}{13.2}{\rmdefault}{\mddefault}{\updefault}{\color[rgb]{0,0,0}$3$}%
}}}}
\put(4996,-1096){\makebox(0,0)[lb]{\smash{{\SetFigFont{11}{13.2}{\rmdefault}{\mddefault}{\updefault}{\color[rgb]{0,0,0}$4$}%
}}}}
\put(3781,-826){\makebox(0,0)[lb]{\smash{{\SetFigFont{11}{13.2}{\rmdefault}{\mddefault}{\updefault}{\color[rgb]{0,0,0}$2$}%
}}}}
\put(3466,-1456){\makebox(0,0)[lb]{\smash{{\SetFigFont{11}{13.2}{\rmdefault}{\mddefault}{\updefault}{\color[rgb]{0,0,0}$1$}%
}}}}
\put(3781,-2086){\makebox(0,0)[lb]{\smash{{\SetFigFont{11}{13.2}{\rmdefault}{\mddefault}{\updefault}{\color[rgb]{0,0,0}$n$}%
}}}}
\put(4996,-1861){\makebox(0,0)[lb]{\smash{{\SetFigFont{11}{13.2}{\rmdefault}{\mddefault}{\updefault}{\color[rgb]{0,0,0}$5$}%
}}}}
\end{picture}%

%% file: contrex.pdf_t
\begin{picture}(0,0)%
\includegraphics{contrex.pdf}%
\end{picture}%
\setlength{\unitlength}{4144sp}%
\begingroup\makeatletter\ifx\SetFigFont\undefined%
\gdef\SetFigFont#1#2#3#4#5{%
  \reset@font\fontsize{#1}{#2pt}%
  \fontfamily{#3}\fontseries{#4}\fontshape{#5}%
  \selectfont}%
\fi\endgroup%
\begin{picture}(3943,3166)(1111,-2454)
\put(1126,-1051){\makebox(0,0)[lb]{\smash{{\SetFigFont{12}{14.4}{\rmdefault}{\mddefault}{\updefault}{\color[rgb]{0,0,0}\colorbox{black!5}{$+$}}%
}}}}
\put(2071,-1906){\makebox(0,0)[lb]{\smash{{\SetFigFont{12}{14.4}{\rmdefault}{\mddefault}{\updefault}{\color[rgb]{0,0,0}\colorbox{black!5}{$+$}}%
}}}}
\put(3781,-2041){\makebox(0,0)[lb]{\smash{{\SetFigFont{12}{14.4}{\rmdefault}{\mddefault}{\updefault}{\color[rgb]{0,0,0}\colorbox{black!5}{$+$}}%
}}}}
\put(2071, 29){\makebox(0,0)[lb]{\smash{{\SetFigFont{12}{14.4}{\rmdefault}{\mddefault}{\updefault}{\color[rgb]{0,0,0}\colorbox{black!5}{$+$}}%
}}}}
\put(3466,-1546){\makebox(0,0)[lb]{\smash{{\SetFigFont{12}{14.4}{\rmdefault}{\mddefault}{\updefault}{\color[rgb]{0,0,0}\colorbox{black!5}{$-$}}%
}}}}
\put(3016,-2311){\makebox(0,0)[lb]{\smash{{\SetFigFont{14}{16.8}{\rmdefault}{\mddefault}{\updefault}{\color[rgb]{0,0,0}$3$}%
}}}}
\put(1666,-961){\makebox(0,0)[lb]{\smash{{\SetFigFont{14}{16.8}{\rmdefault}{\mddefault}{\updefault}{\color[rgb]{0,0,0}$0$}%
}}}}
\put(3061,389){\makebox(0,0)[lb]{\smash{{\SetFigFont{14}{16.8}{\rmdefault}{\mddefault}{\updefault}{\color[rgb]{0,0,0}$2$}%
}}}}
\put(4366,-961){\makebox(0,0)[lb]{\smash{{\SetFigFont{14}{16.8}{\rmdefault}{\mddefault}{\updefault}{\color[rgb]{0,0,0}$1$}%
}}}}
\put(4951,-871){\makebox(0,0)[lb]{\smash{{\SetFigFont{12}{14.4}{\rmdefault}{\mddefault}{\updefault}{\color[rgb]{0,0,0}\colorbox{black!5}{$+$}}%
}}}}
\put(4006, 29){\makebox(0,0)[lb]{\smash{{\SetFigFont{12}{14.4}{\rmdefault}{\mddefault}{\updefault}{\color[rgb]{0,0,0}\colorbox{black!5}{$+$}}%
}}}}
\put(3016,-601){\makebox(0,0)[lb]{\smash{{\SetFigFont{12}{14.4}{\rmdefault}{\mddefault}{\updefault}{\color[rgb]{0,0,0}\colorbox{black!5}{$-$}}%
}}}}
\put(2926,-1231){\makebox(0,0)[lb]{\smash{{\SetFigFont{12}{14.4}{\rmdefault}{\mddefault}{\updefault}{\color[rgb]{0,0,0}\colorbox{black!5}{$-$}}%
}}}}
\put(2476,-376){\makebox(0,0)[lb]{\smash{{\SetFigFont{12}{14.4}{\rmdefault}{\mddefault}{\updefault}{\color[rgb]{0,0,0}\colorbox{black!5}{$-$}}%
}}}}
\end{picture}%

%% file: NMcodeN.pdf_t
\begin{picture}(0,0)%
\includegraphics{NMcodeN.pdf}%
\end{picture}%
\setlength{\unitlength}{4144sp}%
\begingroup\makeatletter\ifx\SetFigFont\undefined%
\gdef\SetFigFont#1#2#3#4#5{%
  \reset@font\fontsize{#1}{#2pt}%
  \fontfamily{#3}\fontseries{#4}\fontshape{#5}%
  \selectfont}%
\fi\endgroup%
\begin{picture}(1599,1959)(-1631,-4933)
\put(-928,-4358){\makebox(0,0)[lb]{\smash{{\SetFigFont{8}{9.6}{\rmdefault}{\mddefault}{\updefault}{\color[rgb]{0,0,0}\colorbox{black!5}{$+$}}%
}}}}
\put(-404,-4336){\makebox(0,0)[lb]{\smash{{\SetFigFont{8}{9.6}{\rmdefault}{\mddefault}{\updefault}{\color[rgb]{0,0,0}\colorbox{black!5}{$+$}}%
}}}}
\put(-1373,-3248){\makebox(0,0)[lb]{\smash{{\SetFigFont{12}{14.4}{\rmdefault}{\mddefault}{\updefault}{\color[rgb]{0,0,0}$1$}%
}}}}
\put(-1373,-4748){\makebox(0,0)[lb]{\smash{{\SetFigFont{12}{14.4}{\rmdefault}{\mddefault}{\updefault}{\color[rgb]{0,0,0}$2$}%
}}}}
\put(-335,-4741){\makebox(0,0)[lb]{\smash{{\SetFigFont{12}{14.4}{\rmdefault}{\mddefault}{\updefault}{\color[rgb]{0,0,0}$3$}%
}}}}
\put(-335,-3948){\makebox(0,0)[lb]{\smash{{\SetFigFont{12}{14.4}{\rmdefault}{\mddefault}{\updefault}{\color[rgb]{0,0,0}$4$}%
}}}}
\end{picture}%

%% file: NMcodeM.pdf_t
\begin{picture}(0,0)%
\includegraphics{NMcodeM.pdf}%
\end{picture}%
\setlength{\unitlength}{4144sp}%
\begingroup\makeatletter\ifx\SetFigFont\undefined%
\gdef\SetFigFont#1#2#3#4#5{%
  \reset@font\fontsize{#1}{#2pt}%
  \fontfamily{#3}\fontseries{#4}\fontshape{#5}%
  \selectfont}%
\fi\endgroup%
\begin{picture}(1599,1959)(349,-4933)
\put(1576,-4336){\makebox(0,0)[lb]{\smash{{\SetFigFont{8}{9.6}{\rmdefault}{\mddefault}{\updefault}{\color[rgb]{0,0,0}\colorbox{black!5}{$+$}}%
}}}}
\put(544,-4009){\makebox(0,0)[lb]{\smash{{\SetFigFont{8}{9.6}{\rmdefault}{\mddefault}{\updefault}{\color[rgb]{0,0,0}\colorbox{black!5}{$-$}}%
}}}}
\put(1147,-3628){\makebox(0,0)[lb]{\smash{{\SetFigFont{8}{9.6}{\rmdefault}{\mddefault}{\updefault}{\color[rgb]{0,0,0}\colorbox{black!5}{$+$}}%
}}}}
\put(1052,-4358){\makebox(0,0)[lb]{\smash{{\SetFigFont{8}{9.6}{\rmdefault}{\mddefault}{\updefault}{\color[rgb]{0,0,0}\colorbox{black!5}{$+$}}%
}}}}
\put(988,-3850){\makebox(0,0)[lb]{\smash{{\SetFigFont{8}{9.6}{\rmdefault}{\mddefault}{\updefault}{\color[rgb]{0,0,0}\colorbox{black!5}{$-$}}%
}}}}
\put(608,-3248){\makebox(0,0)[lb]{\smash{{\SetFigFont{12}{14.4}{\rmdefault}{\mddefault}{\updefault}{\color[rgb]{0,0,0}$1$}%
}}}}
\put(608,-4748){\makebox(0,0)[lb]{\smash{{\SetFigFont{12}{14.4}{\rmdefault}{\mddefault}{\updefault}{\color[rgb]{0,0,0}$2$}%
}}}}
\put(1645,-4741){\makebox(0,0)[lb]{\smash{{\SetFigFont{12}{14.4}{\rmdefault}{\mddefault}{\updefault}{\color[rgb]{0,0,0}$3$}%
}}}}
\put(1645,-3948){\makebox(0,0)[lb]{\smash{{\SetFigFont{12}{14.4}{\rmdefault}{\mddefault}{\updefault}{\color[rgb]{0,0,0}$4$}%
}}}}
\end{picture}%

%% file: figNM1.pdf_t
\begin{picture}(0,0)%
\includegraphics{figNM1.pdf}%
\end{picture}%
\setlength{\unitlength}{4144sp}%
\begingroup\makeatletter\ifx\SetFigFont\undefined%
\gdef\SetFigFont#1#2#3#4#5{%
  \reset@font\fontsize{#1}{#2pt}%
  \fontfamily{#3}\fontseries{#4}\fontshape{#5}%
  \selectfont}%
\fi\endgroup%
\begin{picture}(2634,3219)(3544,-8848)
\put(4186,-7396){\makebox(0,0)[lb]{\smash{{\SetFigFont{12}{14.4}{\rmdefault}{\mddefault}{\updefault}{\color[rgb]{0,0,0}\colorbox{black!5}{\color{gray}$+$}}%
}}}}
\put(4006,-7396){\makebox(0,0)[lb]{\smash{{\SetFigFont{12}{14.4}{\rmdefault}{\mddefault}{\updefault}{\color[rgb]{0,0,0}\colorbox{black!5}{{\color{gray}$-$}}}%
}}}}
\put(4771,-7846){\makebox(0,0)[lb]{\smash{{\SetFigFont{12}{14.4}{\rmdefault}{\mddefault}{\updefault}{\color[rgb]{0,0,0}\colorbox{black!5}{$+$}}%
}}}}
\put(4141,-8400){\makebox(0,0)[lb]{\smash{{\SetFigFont{17}{20.4}{\rmdefault}{\mddefault}{\updefault}{\color[rgb]{0,0,0}$j$}%
}}}}
\put(4141,-6271){\makebox(0,0)[lb]{\smash{{\SetFigFont{17}{20.4}{\rmdefault}{\mddefault}{\updefault}{\color[rgb]{0,0,0}$i$}%
}}}}
\put(5613,-7264){\makebox(0,0)[lb]{\smash{{\SetFigFont{17}{20.4}{\rmdefault}{\mddefault}{\updefault}{\color[rgb]{0,0,0}$k$}%
}}}}
\end{picture}%

%% file: figNM1b.pdf_t
\begin{picture}(0,0)%
\includegraphics{figNM1b.pdf}%
\end{picture}%
\setlength{\unitlength}{4144sp}%
\begingroup\makeatletter\ifx\SetFigFont\undefined%
\gdef\SetFigFont#1#2#3#4#5{%
  \reset@font\fontsize{#1}{#2pt}%
  \fontfamily{#3}\fontseries{#4}\fontshape{#5}%
  \selectfont}%
\fi\endgroup%
\begin{picture}(2634,3219)(6874,-8848)
\put(8101,-7846){\makebox(0,0)[lb]{\smash{{\SetFigFont{12}{14.4}{\rmdefault}{\mddefault}{\updefault}{\color[rgb]{0,0,0}\colorbox{black!5}{$+$}}%
}}}}
\put(7471,-8400){\makebox(0,0)[lb]{\smash{{\SetFigFont{17}{20.4}{\rmdefault}{\mddefault}{\updefault}{\color[rgb]{0,0,0}$j$}%
}}}}
\put(8943,-7264){\makebox(0,0)[lb]{\smash{{\SetFigFont{17}{20.4}{\rmdefault}{\mddefault}{\updefault}{\color[rgb]{0,0,0}$k$}%
}}}}
\put(8943,-7264){\makebox(0,0)[lb]{\smash{{\SetFigFont{17}{20.4}{\rmdefault}{\mddefault}{\updefault}{\color[rgb]{0,0,0}$k$}%
}}}}
\put(7471,-6271){\makebox(0,0)[lb]{\smash{{\SetFigFont{17}{20.4}{\rmdefault}{\mddefault}{\updefault}{\color[rgb]{0,0,0}$i$}%
}}}}
\end{picture}%